\documentclass[showpacs,superscriptaddress,amsmath,amssymb]{revtex4}
\usepackage{graphicx}
\usepackage{bm}

\begin{document}

\title{On soliton structure of higher order (2+1)-dimensional equations of a relaxing medium beneath high-frequency perturbations}

\author{Kuetche Kamgang Victor}
\email{vkuetche@yahoo.fr}

\affiliation{Department of Physics, Faculty of Science, University
of Yaounde I, P.O. Box. 812, Cameroon}

\author{Bouetou Bouetou Thomas}
\email{tbouetou@yahoo.fr}

\affiliation{Ecole Nationale Sup$\acute{e}$rieure Polytechnique,
University of Yaounde I, P.O. Box. 8390, Cameroon}

\affiliation{The Abdus Salam International Centre for Theoretical
Physics, P.O. Box 586, Strada Costiera, II-34014, Trieste, Italy}

\author{Timoleon Crepin Kofane}
\email{tckofane@yahoo.com}

\affiliation{Department of Physics, Faculty of Science, University
of Yaounde I, P.O. Box. 812, Cameroon}

\affiliation{The Abdus Salam International Centre for Theoretical
Physics, P.O. Box 586, Strada Costiera, II-34014, Trieste, Italy}

\date{\today}

\begin{abstract}
We investigate the soliton structure of novel (2+1)-dimensional
nonlinear partial differential evolution (NLPDE) equations which
may govern the behavior of a barothropic relaxing medium beneath
high-frequency perturbations. As a result, we may derive some
soliton solutions amongst which three typical pattern formations
with loop-, cusp- and hump-like shapes.
\end{abstract}

\pacs{02.30.Ik, 05.45.Yv}

%\keywords{2+1)-dimensional nonlinear partial differential evolution,barothropic relaxing
% medium, soliton solutions}

\maketitle

In real natural world, there exist very complicated phenomena
closely related to nonlinear systems. However, in various cases,
the real natural phenomena are too intricate to be described only
by virtue of (1+1)-dimensional NLPDE equations. Thus, a great deal
of interest has recently been paid to higher dimensional NLPDE
equations, especially, the (2+1)-dimensional cases (\cite{Tang}
and references therein). Such higher dimensional systems have been
shown to possess soliton solutions which may arise as a balance
between nonlinearity and dispersion. One underlying query that may
arise, consists of the survival of the soliton properties of
solutions to such systems with higher order nonlinearity. In the
wake of such query, we consider a barothropic relaxing medium
$p=p(\rho,\lambda)$ under high-frequency perturbations $p'\ll
p_{0}$. The quantities $p$ and $\rho$ may stand for pressure and
mass density, respectively. The constant $\lambda$ is an
additional parameter. The pressure $p_{0}$ is measured at the
unperturbed state. For $\lambda=1$, $p=p_{f}$, inner interaction
processes are frozen and the velocities for fast processes may be
defined as $v_{f}^{2}=\frac{dp_{f}}{d\rho}$. For $\lambda=0$,
$p=p_{e}$, there is local thermodynamic equilibrium and the
velocities for slow processes may be defined as
$v_{e}^{2}=\frac{dp_{e}}{d\rho}$.

A few years ago, Danylenko et al. \cite{Dan} have proposed
dynamical equations of state for multicomponent relaxing media. In
the wake of such result, after a suitable expansion of the
specific volume $V\equiv\rho^{-1}$ as power series of perturbation
$p'$ with accuracy $o(p'^{3})$, the following dynamical equation
may be written down
\begin{eqnarray}
\tau\partial_{t}\left(\partial^{2}_{x_{m}x^{m}}p'-\frac{1}{v_{f}^{2}}\partial_{t}^{2}p'
+\alpha_{f}\partial_{t}^{2}p'^{2}+a_{f}\partial_{t}^{2}p'^{3}\right)+\partial^{2}_{x_{m}x^{m}}p'-\frac{1}{v_{e}^{2}}\partial_{t}^{2}p'
+\alpha_{e}\partial_{t}^{2}p'^{2}+a_{e}\partial_{t}^{2}p'^{3}=0,\quad
(m=1,\cdots,N),\label{eq1}
\end{eqnarray}
where $\tau$ is the relaxation time; constants $\alpha_{e}$ and
$\alpha_{f}$ represent positive-valued second order expansion
c$\oe$fficients of $V_{e}$ and $V_{f}$, respectively; constants
$a_{e}$ and $a_{f}$ stand for positive-valued third order
expansion c$\oe$fficients of $V_{e}$ and $V_{f}$, respectively;
quantity $x^{m}\equiv (x,y,z,\cdots)$ may stand for
position-vector and
$\partial^{2}_{x_{m}x^{m}}=\partial^{2}_{x_{1}x^{1}}+\partial^{2}_{x_{2}x^{2}}+\partial^{2}_{x_{3}x^{3}}+\cdots$,
$(m=1,\cdots,N)$. It is noted that any repeated index refers to
summation with respect to Einstein's notation. In order to
investigate the eq. (\ref{eq1}), the multiscale method \cite{Nay,
Nit} may be useful. Thus, defining the quantity
$\epsilon=\tau\omega$ ($\omega$ being the frequency of the
processes) chosen to be small (large) parameter, after introducing
the following independent variables $T_{0}=t\omega$,\quad
$T_{-4}=t\omega\epsilon^{-4}$,\quad $X_{0}^{m}=x^{m}\omega$,
$X_{-4}^{m}=x^{m}\omega\epsilon^{-4}$, $(m=1,\cdots,N)$ into eq.
(\ref{eq1}), seven coupled equations may be derived. From this
coupled system, there may be two leading equations expressed in
terms of $T_{0}$ and $X_{0}^{m}$ $(m=1,\cdots,N)$, describing
low-frequency perturbations, and the two other expressed in terms
of $T_{-4}$ and $X_{-4}^{m}$ $(m=1,\cdots,N)$ describing
high-frequency perturbations. Thus, focusing our interest only to
high frequency perturbations, we may derive the following
evolution equation
\begin{eqnarray}
(\partial_{x_{n}}+\partial_{x_{m}})(\partial_{x_{n}}+\partial_{x_{m}})\left(p'+\alpha_{f}v_{f}^{2}p'^{2}+a_{f}v_{f}^{2}p'^{3}\right)-v_{f}^{-2}\partial_{tt}^{2}p'+\beta_{f}\partial_{x_{m}}J^{m}+\gamma_{f}p'=0,\quad
(n<m),\label{eq2}
\end{eqnarray}
where $J^{m}=p'$ $(m=1,\cdots,N)$, and the quantities $\beta_{f}$
and $\gamma_{f}$ may be expressed as follows
\begin{eqnarray}
\beta_{f}=\frac{v_{f}^{2}-v_{e}^{2}}{\tau v_{e}^{2}v_{f}},\quad
\gamma_{f}=\frac{v_{f}^{4}-v_{e}^{4}}{2\tau^{2}v_{e}^{4}v^{2}_{f}}.\label{eq3}
\end{eqnarray}
 Eq. (\ref{eq2}) may be obtained in the following way. A
 dispersion relation for the linearized eq. (\ref{eq2}) may be written down in the
 form
 $v_{f}^{-2}\omega^{2}=\left(k_{n}+k_{m}\right)\left(k_{n}+k_{m}\right)+j\beta_{f}\sum_{m=1}^{N}k_{m}-\gamma_{f}$ with $(n<
 m=1,\cdots,N)$. The nonlinear terms may be reconstructed in agreement with the
 initial equation. We may note here that eq. (\ref{eq2}) has
 dissipative $\beta_{f}$-terms and dispersive $\gamma_{f}$-terms.
 If $\alpha_{f}=a_{f}=\beta_{f}=0$, eq. (\ref{eq2}) reduces to a
 typical modified Klein-Gordon equation in (N+1)-dimensional
 space. We particularly focus our interests to $N=2$ case. Then,
 it comes
 \begin{eqnarray}
(\partial_{x}+\partial_{y})^{2}p'-v_{f}^{-2}\partial_{tt}^{2}p'+\alpha_{f}v_{f}^{2}(\partial_{x}+\partial_{y})^{2}p'^{2}+a_{f}v_{f}^{2}(\partial_{x}+\partial_{y})^{2}p'^{3}+\beta_{f}(\partial_{x}+\partial_{y})p'+\gamma_{f}p'=0.
\label{eq4}
\end{eqnarray}
If $a_{f}=0$ and y-terms removed, eq. (\ref{eq4}) may reduce to
the (1+1)-dimensional Vakhnenko (V) equation
\cite{Vak,Vak1,Vak2,Vak3,Vak4}. We need the following accuracy
\begin{eqnarray}
(\partial_{x}+\partial_{y})^{2}-v_{f}^{-2}\partial_{tt}^{2}\approx
2(\partial_{x}+\partial_{y})\left(\partial_{x}+\partial_{y}+v_{f}^{-1}\partial_{t}\right),\label{eq5}
\end{eqnarray}
to further investigate the eq. (\ref{eq4}). Thus, it may be
interesting to consider two important cases: $\alpha_{f}=0$ and
$\alpha_{f}\neq 0$.

\begin{enumerate}
    \item Case $\alpha_{f}=0$,

then, eq. (\ref{eq4}) may be transformed to
\begin{eqnarray}
(\partial_{\tilde{x}}+\partial_{\tilde{y}})\left[\partial_{\tilde{t}}-\frac{1}{2}\tilde{u}^{2}(\partial_{\tilde{x}}+\partial_{\tilde{y}})\right]\tilde{u}+\tilde{\alpha}
(\partial_{\tilde{x}}+\partial_{\tilde{y}})\tilde{u}-\tilde{u}=0,\label{eq6}
\end{eqnarray}
provided
\begin{eqnarray}
\tilde{x}=-\sqrt{\frac{\gamma_{f}}{6}}(x-v_{f}t),\quad
\tilde{y}=-\sqrt{\frac{\gamma_{f}}{6}}(y-v_{f}t),\quad
\tilde{t}=v_{f}\sqrt{\frac{3\gamma_{f}}{2}}t,\quad
\tilde{u}=\alpha_{f}v_{f}^{2}p',\quad
\tilde{\alpha}=\frac{\beta_{f}}{\sqrt{6\gamma_{f}}},\label{eq7}
\end{eqnarray}
hold. Performing variable transformation, we introduce new
independent variables $X$, $T_{1}$ and $T_{2}$ as follows
\begin{eqnarray}
\tilde{x}=T_{1}-\frac{1}{2}\int_{-\infty}^{X}U^{2}dX'+\tilde{x}_{0},\quad
\tilde{y}=T_{2}-\frac{1}{2}\int_{-\infty}^{X}U^{2}dX'+\tilde{y}_{0},\quad
\tilde{t}=X,\label{eq8}
\end{eqnarray}
where $\tilde{x}_{0}$ and $\tilde{y}_{0}$ stand for arbitrary
constants and
$\tilde{u}(\tilde{x},\tilde{y},\tilde{t})=U(T_{1},T_{2},X)$. Eq.
(\ref{eq6}) may be rewritten as follows
\begin{eqnarray}
U_{XT_{1}}+U_{XT_{2}}+\tilde{\alpha}(U_{T_{1}}+U_{T_{2}})-(1+\varphi+\phi)U=0,\label{eq9}
\end{eqnarray}
where $\varphi=-\int_{-\infty}^{X}UU_{T_{1}}dX'$ and
$\phi=-\int_{-\infty}^{X}UU_{T_{2}}dX'$. Subscripts with respect
to $X$, $T_{1}$ and $T_{2}$ may denote partial differentiations.
Using the ansates $\varphi=Z_{1T_{1}}-1$ and $\phi=Z_{2T_{2}}-1$,
eq. (\ref{eq6}) may bilinearize to
\begin{eqnarray}
(D_{X}D_{T_{1}}+D_{X}D_{T_{2}}+\tilde{\alpha}(D_{T_{1}}+D_{T_{2}})-1)G\cdot
F=0,\quad D_{X}^{2}F\cdot F=\frac{1}{2}G^{2},\label{eq10}
\end{eqnarray}
provided
\begin{eqnarray}
U=\frac{G}{F},\quad Z_{1}=T_{1}-2\left(\ln(F)\right)_{X},\quad
Z_{2}=T_{2}-2\left(\ln(F)\right)_{X},\label{eq11}
\end{eqnarray}
hold. Notations $D_{X}$, $D_{T_{1}}$ and $D_{T_{2}}$ denote Hirota
operators \cite{Hir1,Hir2}. Expanding $F$ and $G$ in a suitable
formal power series, a one-soliton solution to eq. (\ref{eq9}) may
be given by
\begin{eqnarray}
U=2Ksech(\theta),\quad
Z_{1}=T_{1}-2K\left[\tanh(\theta)+1\right],\quad
Z_{2}=T_{2}-2K\left[\tanh(\theta)+1\right].\label{eq12}
\end{eqnarray}
where $\theta=KX-\omega_{1}T_{1}-\omega_{1}T_{1}+\theta_{0}$,
$\theta_{0}$ being an arbitrary constant. The dispersion relation
may be given by
\begin{eqnarray}
(K+\tilde{\alpha})(\omega_{1}+\omega_{2})+1=0.\label{eq13}
\end{eqnarray}
As a result, assuming that $\omega_{1}=Kv_{1}$ and
$\omega_{2}=Kv_{2}$ such that $v=v_{1}+v_{2}<0$, we may find out
three typical pattern formations. Indeed,
\begin{enumerate}
    \item for
$\tilde{\alpha}<\sqrt{\frac{1}{2|v|}}$, loop-like pattern may be
obtained (see FIG. 1 in the case of $v=-0.24$);
    \item for
$\tilde{\alpha}=\tilde{\alpha}_{c}=\sqrt{\frac{1}{2|v|}}$,
cusp-like pattern may be obtained (see FIG. 2 in the case of
$v=-0.24$ where
$\tilde{\alpha}_{c}=1,4433756729740644112728719512549$);

\item finally, for $\tilde{\alpha}>\sqrt{\frac{1}{2|v|}}$,
hump-like pattern may be obtained (see FIG. 3 in the case of
$v=-0.24$).
\end{enumerate}
These illustrative curves are plotted at initial time
$\tilde{t}=0$. An essential remark on eq. (\ref{eq6}) should be
noted here. In the wake of the results got from ref. \cite{Kuek},
an extended complex-valued form of eq. (\ref{eq6}) may be found as
follows
\begin{eqnarray}
(\partial_{\tilde{x}}+\partial_{\tilde{y}})\left[\partial_{\tilde{t}}-\frac{1}{2}\tilde{q}\tilde{q}^{\star}(\partial_{\tilde{x}}+\partial_{\tilde{y}})\right]\tilde{q}+\tilde{\alpha}
(\partial_{\tilde{x}}+\partial_{\tilde{y}})\tilde{q}-\tilde{q}=0,\label{eq14}
\end{eqnarray}
where $q=q^{r}+\imath q^{im}$ may stand for a complex-valued
observable, symbol $(\star)$ may refer to complex conjugation and
$\imath^{2}=-1$. Thus, one may derive the following set
\begin{subequations}
\label{eq15}
\begin{eqnarray}
Q^{r}_{XT_{1}}+Q^{r}_{XT_{2}}+\tilde{\alpha}\left(Q^{r}_{T_{1}}+Q^{r}_{T_{2}}\right)-\left(1+Z_{1T_{1}}+Z_{2T_{2}}\right)Q^{r}=0,\label{eq15a}
\end{eqnarray}
\begin{eqnarray}
Q^{im}_{XT_{1}}+Q^{im}_{XT_{2}}+\tilde{\alpha}\left(Q^{im}_{T_{1}}+Q^{im}_{T_{2}}\right)-\left(1+Z_{1T_{1}}+Z_{2T_{2}}\right)Q^{im}=0,\label{eq15b}
\end{eqnarray}
\begin{eqnarray}
Z_{1XT_{1}}=-\left(Q^{r}Q^{r}_{T_{1}}+Q^{im}Q^{im}_{T_{1}}\right),\quad
Z_{1XT_{2}}=-\left(Q^{r}Q^{r}_{T_{2}}+Q^{im}Q^{im}_{T_{2}}\right),\label{eq15c}
\end{eqnarray}
\end{subequations}
where $\tilde{q}(\tilde{x},\tilde{y},\tilde{t})=Q(T_{1},T_{2},X)$
such that independent variables $\tilde{x}$, $\tilde{y}$ and
$\tilde{t}$ now stand for
\begin{eqnarray}
\tilde{x}=T_{1}-\frac{1}{2}\int_{-\infty}^{X}QQ^{\star}dX'+\tilde{x}_{0},\quad
\tilde{y}=T_{2}-\frac{1}{2}\int_{-\infty}^{X}QQ^{\star}dX'+\tilde{y}_{0},\quad
\tilde{t}=X.\label{eq16}
\end{eqnarray}
It comes that eq. (\ref{eq15}) may be closely related to the
coupled dispersionless systems \cite{Ala,Kuet} recently
investigated by Kakuhata and Konno \cite{Kak1,Kak2}. Besides, eq.
(\ref{eq15}) may be also observed as a coupled (2+1)-dimensional
version of the complex-form of the Sch$\ddot{a}$fer-Wayne short
pulse (SWSP) equation \cite{Sch} that has been subject to many
recent investigations \cite{Kuek,Par,Kuek1,Sak,Kuek2}. A
one-soliton solution to eq. (\ref{eq15}) may be written as follows
\begin{eqnarray}
Q=A\exp(\theta^{im})sech(\theta^{r}),\quad
Z_{1}=T_{1}-2K^{r}\left[\tanh(\theta^{r})+1\right],\quad
Z_{2}=T_{2}-2K^{r}\left[\tanh(\theta^{r})+1\right],\label{eq17}
\end{eqnarray}
where
$\theta=\theta^{r}+\imath\theta^{im}=KX-T_{1}\omega_{1}-T_{2}\omega_{2}+\theta_{0}$,
$\theta_{0}$ being an arbitrary complex-valued constant,
$K=K^{r}+\imath K^{im}$ and
$\omega_{j}=\omega_{j}^{r}+\imath\omega_{j}^{im}$ $(j=1,2)$. The
amplitude $A$ may be given by $A=2K^{r}$. The corresponding
dispersion relation may be given by eq. (\ref{eq13}). As a result,
one may easily find that solutions given by eq. (\ref{eq17}) in
terms of $Q^{r}$ and $Q^{im}$, possess a nonzero angular momentum
i.e. the previous patterns depicted above may rotate with an
angular frequency $\Omega=\omega_{1}^{im}+\omega_{2}^{im}$.
Concretely, if we assume that $\omega_{1}^{r}=v_{1}K^{r}$,
$K^{im}=\zeta \omega_{1}^{im}$, $\omega_{2}^{r}=v_{2}K^{r}$,
$K^{im}=\zeta \omega_{2}^{im}$ and $v=v_{1}+v_{2}<0$, $K^{r}$ and
$\Omega$ may be expressed as follows
\begin{eqnarray}
K^{r}=\frac{\tilde{\alpha}}{\zeta
|v|-1},\quad\Omega=\frac{1}{\sqrt{\zeta}}
\sqrt{1-\frac{\zeta\tilde{\alpha}^{2}v^{2}}{(\zeta
v-1)^{2}}},\label{eq17a}
\end{eqnarray}
provided $\frac{1}{|v|}<\zeta<\frac{2}{|v|}$. Thus, setting
$\alpha_{c}=\frac{\zeta |v|-1}{\sqrt{2|v|}}$ and
$\tilde{\alpha}_{s}=\frac{\zeta |v|-1}{|v|\sqrt{\zeta}}$, it comes
that,
\begin{enumerate}
    \item for $\alpha\geq\tilde{\alpha}_{s}$, no rotating
pattern formation is expected;
    \item for $\tilde{\alpha}_{c}<\alpha<\tilde{\alpha}_{s}$, rotating loop-like
pattern formation is derived (see FIG.1 for $v=-0.24$ and
$\zeta=13/3$);
    \item for $\tilde{\alpha}=\alpha_{c}$, rotating
cusp-like pattern formation is derived (see FIG.2 for $v=-0.24$
and $\zeta=13/3$ where
$\alpha_{c}=0,057735026918962576450914878050196$);
    \item finally, for
$\alpha<\tilde{\alpha}_{c}$, rotating hump-like pattern formation
is derived (see FIG.3 for $v=-0.24$ and $\zeta=13/3$).
\end{enumerate}

\item Case $\alpha_{f}\neq 0$,

then, eq. (\ref{eq4}) may lead to
\begin{eqnarray}
(\partial_{\tilde{x}}+\partial_{\tilde{y}})\left[\partial_{\tilde{t}}+\tilde{u}(\partial_{\tilde{x}}+\partial_{\tilde{y}})+\frac{1}{2}\tilde{u}^{2}(\partial_{\tilde{x}}+\partial_{\tilde{y}})\right]\tilde{u}+\tilde{\alpha}
(\partial_{\tilde{x}}+\partial_{\tilde{y}})\tilde{u}+\tilde{u}=0,\label{eq18}
\end{eqnarray}
provided
\begin{eqnarray}
\tilde{x}=\sqrt{\frac{3a_{f}\gamma_{f}}{2\alpha_{f}^{2}v_{f}^{2}}}(x-v_{f}t),\quad
\tilde{y}=\sqrt{\frac{3a_{f}\gamma_{f}}{2\alpha_{f}^{2}v_{f}^{2}}}(y-v_{f}t),\quad
\tilde{t}=\sqrt{\frac{\gamma_{f}}{6a_{f}}}\alpha_{f}v_{f}^{2}t,\quad
\tilde{u}=\frac{3a_{f}}{\alpha_{f}}p',\quad
\tilde{\alpha}=\frac{\beta_{f}}{\alpha_{f}v_{f}}\sqrt{\frac{3a_{f}}{2\gamma_{f}}},\label{eq19}
\end{eqnarray}
hold. We introduce new independent variables $X'$, $T'_{1}$ and
$T'_{2}$ as follows
\begin{eqnarray}
\tilde{x}=T'_{1}+\frac{1}{2}\int_{-\infty}^{X'}\left(U^{2}+2U\right)dS+\tilde{x}'_{0},\quad
\tilde{y}=T'_{2}+\frac{1}{2}\int_{-\infty}^{X'}\left(U^{2}+2U\right)dS+\tilde{y}'_{0},\quad
\tilde{t}=X',\label{eq20}
\end{eqnarray}
where $\tilde{x}'_{0}$ and $\tilde{y}'_{0}$ stand for arbitrary
constants and
$\tilde{u}(\tilde{x},\tilde{y},\tilde{t})=U(T'_{1},T'_{2},X')$.
Then eq. (\ref{eq18}) becomes
\begin{eqnarray}
U_{X'T'_{1}}+U_{X'T'_{2}}+\tilde{\alpha}(U_{T'_{1}}+U_{T'_{2}})+(1+\varphi'+\phi')U=0,\label{eq21}
\end{eqnarray}
where $\varphi'=\int_{-\infty}^{X'}(U+1)U_{T_{1}}dS$ and
$\phi'=\int_{-\infty}^{X}(U+1)U_{T_{2}}dS$. Using the ansates
$\varphi'=-Z_{1T'_{1}}+1$ and $\phi'=-Z_{2T'_{2}}+1$, eq.
(\ref{eq18}) may be bilinearized as
\begin{eqnarray}
(D_{X'}D_{T'_{1}}+D_{X'}D_{T'_{2}}+\tilde{\alpha}(D_{T'_{1}}+D_{T'_{2}})+1)G\cdot
F=0,\quad D_{X}^{2}F\cdot
F=\frac{1}{2}\left(G^{2}+2GF\right),\label{eq22}
\end{eqnarray}
provided
\begin{eqnarray}
U=\frac{G}{F},\quad Z_{1}=T'_{1}-2\left(\ln(F)\right)_{X'},\quad
Z_{2}=T'_{2}-2\left(\ln(F)\right)_{X'},\label{eq23}
\end{eqnarray}
hold. Thus expanding suitably the functions $G$ and $F$ in power
series according to Hirota method \cite{Hir1,Hir2}, one-soliton
solutions of kink-like pattern may be expected and discussed by
means of the dispersion relation derived from eq. (\ref{eq22}).
\end{enumerate}

In summary, we have given a general model of a (2+1)-dimensional
NLPDE equation (see eq. (\ref{eq14})) with the soliton structure.
This novel (2+1)-dimensional NLPDE equation (eq. (\ref{eq14})) may
be observed as a coupled (2+1)-dimensional version of the
(1+1)-dimensional complex-SWSP equation \cite{Kuek}. This novel
equation with the other one (see eq. (\ref{eq18})) may deserve
further scientific interests both from the viewpoint of the
investigation of the propagation of high-frequency perturbations
and from the viewpoint of the existence of stable wave formations.
We may actually believe that these equations may be valuable for
studies on soliton theory, geodynamics, plasma physics,
hydrodynamics, condensed matter, string theory, nonlinear optics,
just to name a few.

%\newpage
\bibliography{PRLVakSw}
%\newpage
%\begin{figure}
%    \begin{center}
%    \includegraphics[width=8cm]{FIGURE1.eps}
%    \caption{Loop-like pattern formation.}\label{FIG1}
%\end{center}
%\end{figure}
%\begin{figure}
%    \begin{center}
%    \includegraphics[width=8cm]{FIGURE2.eps}
%    \caption{Cusp-like pattern formation.}\label{FIG2}
%    \end{center}
%\end{figure}
%\begin{figure}
%    \begin{center}
%    \includegraphics[width=8cm]{FIGURE3.eps}
%    \caption{Hump-like pattern formation.}\label{FIG3}
%    \end{center}
%\end{figure}
\end{document}